\documentclass[a4paper]{spie}
\usepackage[]{graphicx}
\usepackage[]{wrapfig}
\usepackage[]{amsmath}

\title{Comparison of the atmospheric properties above Dome A, Dome C and the South Pole}

\author{Susanna Hagelin\supit{a,b}, Elena Masciadri\supit{a}, Franck
  Lascaux\supit{a} and Jeff Stoesz\supit{a}
\skiplinehalf
\supit{a}INAF Osservatorio Astrofisico di Arcetri, Largo Enrico
  Fermi 5, I-501 25 Firenze, Italy \\
\supit{b}Department of Earth Sciences, Uppsala
  Universitet, Villav\"agen 16, S-752 36 Uppsala, Sweden
}

\authorinfo{E-mail: hagelin@arcetri.astro.it}

\begin{document}
\maketitle


\begin{abstract}
The atmospheric properties above three sites on the Internal Antarctic
Plateau are investigated for astronomical applications calculating the
monthly median of the analysis-data from ECMWF (European Centre for
Medium-Range Weather Forecasts) for an entire year (2005) thus covering
all seasons. Radiosoundings extended on a yearly time scale from Dome
C and the South Pole are used to verify the reliability of the
analyses in the free atmosphere and to study the wind speed in the
first 100 m as the analysis-data are not optimized for this
altitude-range. The wind speed in the free atmosphere is obtained from
the ECMWF analyses from all three sites. It appears that the strength
of the wind speed in the upper atmosphere in winter is correlated
to the distance of the site from the centre of the polar high.

The Richardson number is employed to investigate the stability of the
free atmosphere and, consequently, the probability to trigger
thermodynamic instabilities above the three sites. We find that, in a
large majority of the cases, the free atmosphere over the Internal
Antarctic Plateau is more stable than at mid-latitude sites.

Given these data we can obtain a ranking of the three sites with
respect to wind speed, in the free atmosphere as well as in the
surface layer, and with respect to the stability of the atmosphere,
using the Richardson number.
\end{abstract}


\section{INTRODUCTION}

The summits of the Internal Antarctic Plateau (Dome A and Dome C) might be among the best places in the world for astronomical
facilities. The free atmosphere is extremely clear with a low humidity
and low levels of turbulence and the optical turbulence appears to be
confined to a narrow surface layer.\cite{Marks99, Law04, Agabi06}. The
height of this surface layer is expected to be lower at the summits
than at the slopes of the plateau. Measurements show that the surface
layer at Dome C\cite{Agabi06} is 36$\pm$10 m while the height measured
at the South Pole\cite{Marks02, Trav03} (situated at a slope) is 220 m
or 270 m.

The largest source of turbulence in the surface layer is the
near-surface winds. The surface winds are triggered by the sloping
terrain in combination with a temperature inversion\cite{Schwe84}. The
inversion is present for a large proportion of the time in Antarctica.
It forms when the incoming solar radiation is less then the
radiative emission from the snow surface which result in a radiative
cooling of the air closest to the surface. The calmest conditions are
found at the summits as the principal cause that triggers the
surface winds, the slope of the terrain, is absent here.

This study is an attempt to better characterize these sites using the
analysis data from the MARS archive, containing the model data from
the GCM (General Circulation Model) of the ECMWF (European Centre for
Medium-Range Weather Forecasts). A detailed description of the
analyses data can be found in Geissler \& Masciadri
(2006).\cite{Ges06} Data has been used for an entire year, 2005 at 00
UTC (unless indicated otherwise), in an attempt to give a
statistical characterization of the differences between the
sites.

The use of data has the huge advantage of being able to access
data from anywhere in the world, and we also have the possibility to
access the past as well as the present and to make simulations of the
future. However a GCM describes the circulation of the whole Earth
and its usually assumed to have a resolution that
\begin{wraptable}[7]{r}{65mm}
\centering
\caption{The locations of the sites}
\medskip
\begin{tabular}{llr@{$^\circ$}l} 
\hline
Site & Lat. & \multicolumn{2}{l}{Long.} \\
\hline
Dome A          & 80$^\circ$22' S &  77 & 21' E \\
Dome C          & 75$^\circ$06' S & 123 & 20' E \\
Dome F          & 77$^\circ$19' S &  39 & 42' E \\
South Pole      & 90$^\circ$00' S &   0 & 00' E \\
\hline
\end{tabular}
\label{ou}  
\end{wraptable}
is too low to
accurately describe the surface layer.\cite{Ges06} Studies have also
appeared\cite{Sadib06} that claims the ECMWF data also are valid, with
a good degree of accuracy, near the surface. This study\cite{Sadib06}
does only treat summer data and it is possible that their conclusion
is not valid for all seasons. A further discussion of this topic is
presented in section 2, which investigates the difference between
ECMWF data and radiosoundings.


In order to study the atmospheric circulation near the surface it is
in principle better to use a mesoscale model as their resolution is
much finer and their physics scheme usually is non-hydrostatic so it
can resolve processes occurring at a smaller scale. However it is well
worth the effort to first investigate the limit to which we can rely
on the global models. It is fundamental to have a clear picture of the
limitations of the ECMWF data as well as to try get the maximum out of
the available information that such a product can give us.

This paper is an attempt to quantify the differences of some critical
meteorological parameters above three sites of astronomical interest,
Dome A, Dome C and the South Pole. (Dome F  will also be discussed with
regards to a few of the parameters.) Measurements are also available from the South Pole and Dome C, whereas no
measurements are yet available from Dome A. 

The scientific goals of this paper are:
\begin{itemize}
\item To make a detailed comparison of the offset/difference between
  radiosoundings and the ECMWF analyses of the wind speed and the
  temperature (the main parameters defining the stability of the
  atmosphere) near the surface during both winter and summer. This
  will permit us to quantify the uncertainty between measurements and
  analyses. The idea is to define the conditions in which the
  ECMWF analyses can be used, with a good level of accuracy, to
  describe the meteorological parameters and then use this tool
  (ECMWF analyses) to characterize a site for which there are no
  measurements.

\item Using data from radiosoundings we will estimate the statistic
median values of the wind speed in the first 150 m at the South Pole
and Dome C, from April to November. With this information we
can quantify which site shows the better characteristics for
astronomical applications.

\item Extending the study of Geissler \& Masciadri (2006)\cite{Ges06} of the
wind speed in the free atmosphere of Dome C also to the South Pole,
Dome A and Dome F. All sites are located on the Internal Antarctic
Plateau but at different latitude and longitude. (Their locations are
indicated in Table \ref{ou}.) In this way we intend to quantify which
site is the best for astronomical applications. The result of this
analysis is fundamental to put in the right perspective the
potentialities of Dome C. As already been discussed by Geissler \&
Masciadri (2006)\cite{Ges06}, in winter the wind speed grows
monotonically above 10 km a.s.l. (above sea level), achieving median
values of the order of $\sim$30 m/s at 25 km a.s.l. Such a strong wind
might trigger an important decrease of the wavefront coherence time
and, as a consequence, the potentiality of these sites might
disappear. Therefore it should be interesting to extend this type of
analysis to more sites on the plateau to retrieve some general
information of the wind speed above the Internal Antarctic Plateau.

\item Extending the analysis of the Richardson number done by Geissler \&
Masciadri (2006)\cite{Ges06} for Dome C to the three sites (South
Pole, Dome C and Dome A) in order to investigate which regions and
periods that are less favourable for the triggering of optical
turbulence and to identify the site with the best characteristics for
astronomical applications. This result should represent the first
estimate of the potentialities of Dome A and we are therefore able to
provide some reliable conclusions about this site even before some
measurements are done on the site. The double interest of this study
is, first and foremost the result in itself and second the opening of a
path to a different approach for a fast and reliable classification of
potential astronomical sites.
\end{itemize}

This contribution is an extract of a more exhaustive analysis by
Hagelin et al.\cite{Hag08}

\section{ECMWF ANALYSES VERSUS RADIOSOUNDINGS}

The ECMWF analyses are compared to radiosoundings from the South Pole
and Dome C in order to investigate the reliability of the
ECMWF analyses over sites on the Internal Antarctic Plateau. The
radiosoundings and the analyses used for this comparison are both from
2006. This year was chosen because of the richer sample of available
radiosoundings. The comparison of analyses data discussed further on
in this paper are from 2005, the reason being that we wish to
investigate a homogeneous data set. In February 2006 the ECMWF changed
the number of vertical levels from 60 to 91.

The median of the difference between ECMWF analyses and radiosoundings
is calculated for winter (June, July and August) as well as
summer (December, January and February). Particular interest is paid
to the first 150 m of the atmosphere as it is important to quantify
the reliability of the analyses data in this layer. In the free
atmosphere there is a noticeable difference between the radiosoundings
from summer and the ones from winter, namely the difference in the
maximum altitude they reach. This fact makes it difficult to study the
reliability of the ECMWF analyses in the high part of the atmosphere
in winter. During this season the balloons frequently explode at
altitudes somewhere around 10 km, probably due to the low
pressure in the high part of the atmosphere in combination with the
very low temperatures. Due to the low pressure the balloon expand and
due to the low temperature, much lower than the average summer
temperature at the same altitude, the material of the balloon is more
fragile and explode easier.

\begin{figure}
\begin{centering}
\includegraphics[scale=0.85]{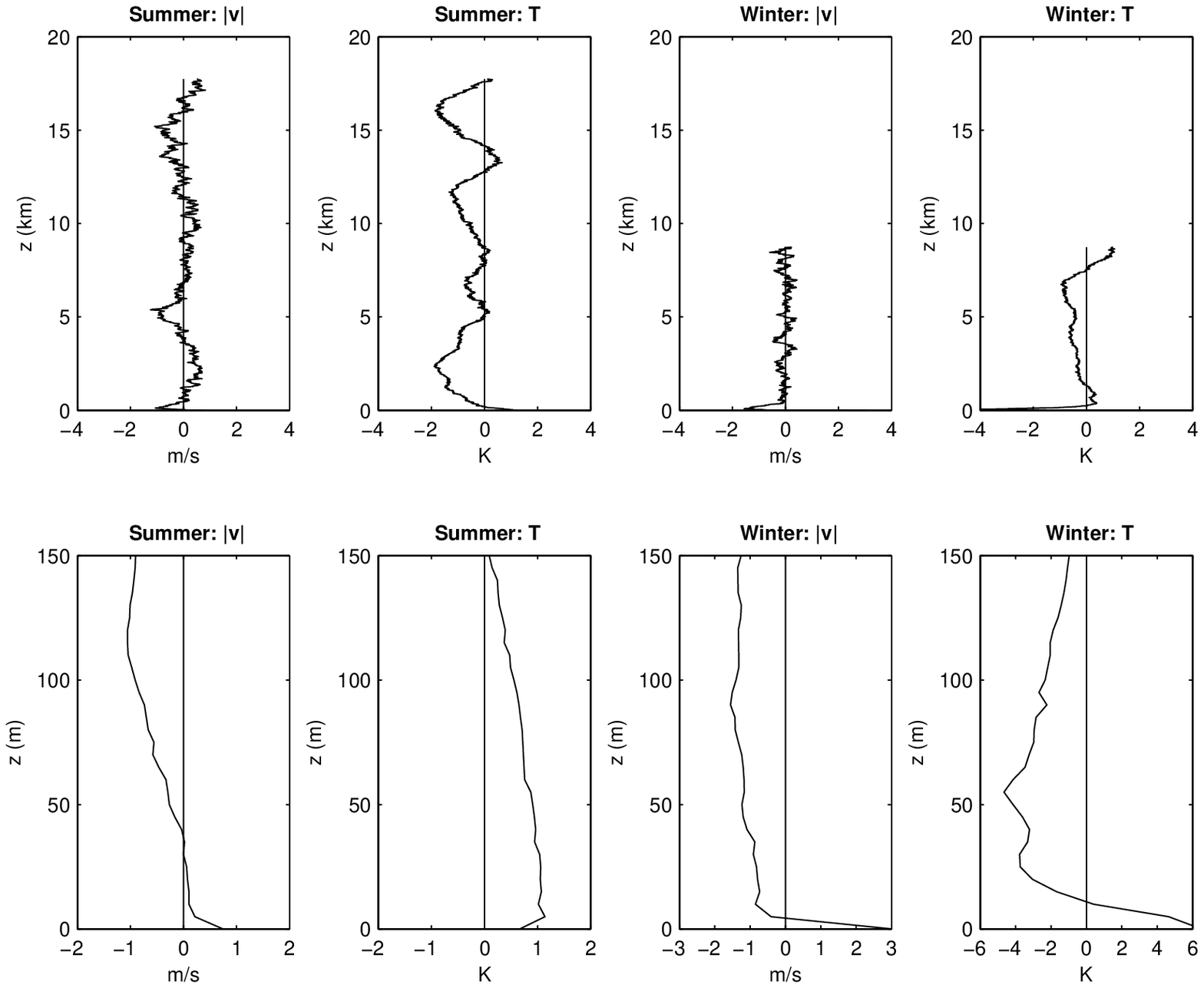}
\caption{The median of the difference of the absolute temperature and
  the wind speed between ECMWF analyses and radiosounding for summer
  (December, January, February) and winter (June, July, August) at
  Dome C in 2006.}
\label{motd}

\includegraphics[scale=0.85]{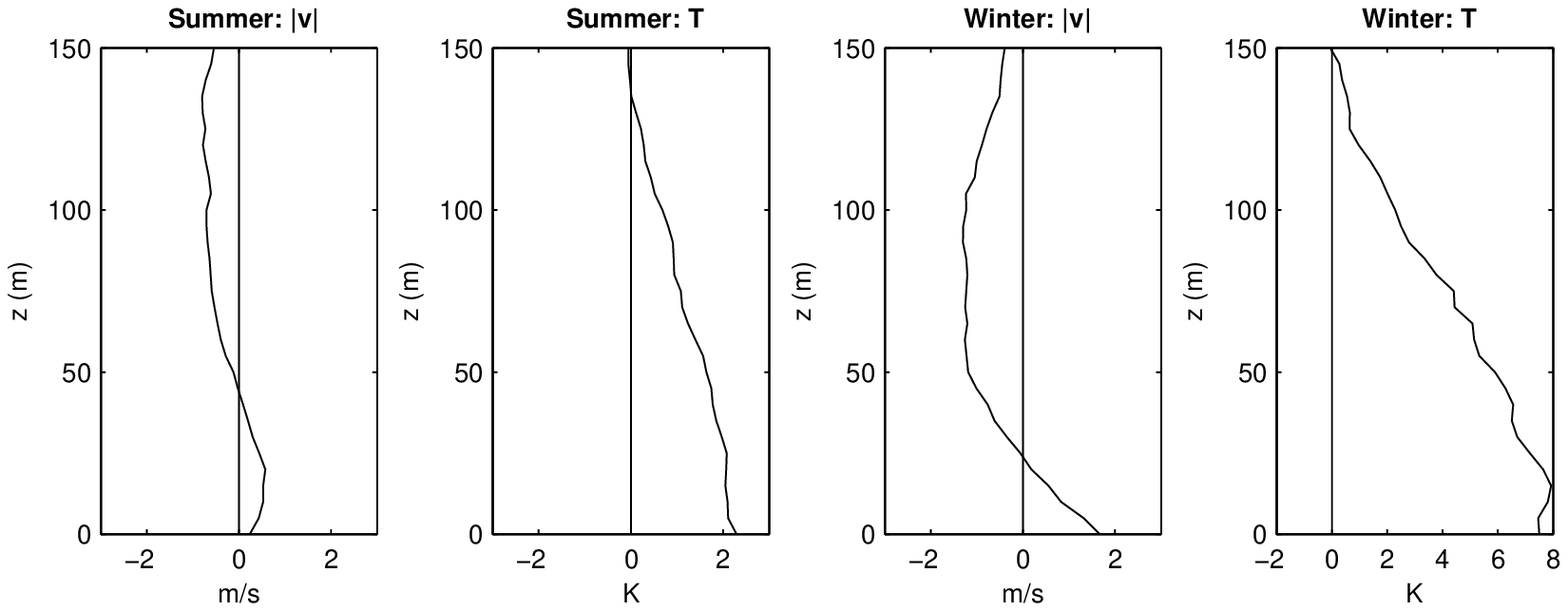}
\caption{The median of the difference of the absolute temperature and
  the wind speed between  ECMWF analyses and radiosoundings for summer
  (December, January, February) and winter (June, July, August) at
  South Pole in 2006.}
\label{motdd}
\end{centering}
\end{figure}

Figure \ref{motd} shows the median of the difference in wind speed and
temperature between ECMWF analyses and radiosoundings for summer and
winter at Dome C. In local summer the median difference in wind speed
never exceeds 1 m/s. Closest to the ground the difference is even
smaller, less than 1 m/s. During local winter the median difference in
wind speed in the free atmosphere never exceeds 0.5 m/s, though the
radiosoundings only reach $\sim$10 km above the ground. The largest
median difference is found near the surface, $\sim$3 m/s during this
season. 

The median difference of the absolute temperature in summer is below 2
K throughout the whole altitude range investigated. Near the surface
this difference is of the order of 1 K, closest to the surface it is
even smaller. During winter in the high part of the atmosphere the
median difference is similar to what is observed in summer, but in the
first 100 m the median difference is significantly larger, more than 6
K nearest the surface.

Figure~\ref{motdd} shows the same output but calculated for the South
Pole. Only the first 150 m are shown, for a comparison in the free
atmosphere we refer the reader to Geissler \& Masciadri
(2006)\cite{Ges06}. Regarding the wind speed the difference at the
South Pole is similar to what is observed at Dome C. The wind speed
difference remain within 1 m/s in the first 150 m in summer. The
analyses show a tendency to overestimate the wind speed near the
ground in winter. The median difference of the absolute temperature is
similar to what is observed at Dome C, $\sim$1 K in summer. Near the
ground the ECMWF-analyses are almost 2 K warmer than the
radiosoundings. The same trend is observed also in winter. However
during this season the analyses are visible much warmer ($\sim$6 K)
than the radiosoundings near the surface.

The statistic uncertainty of the radiosounding data is reported in
Hagelin et al. (2008)\cite{Hag08}. The same paper also report a study
of the comparison of the median value of the wind speed of the
radiosoundings in their first point (the most critical point) in the
central months of the winter and
AWS (Automatic Weather Stations). The values obtained with the
different types of measurements agree in an excellent way. However
this comparison permitted us to put in evidence a small bias ($\sim$2
K) in the radiosoundings. This decreases the overestimation of the
temperature in the ECMWF analyses to 4-5 K.

Summarizing, the wind speed is well reconstructed by the ECMWF
analyses, with exception of the surface layer in winter where the
ECMWF analyses show a tendency of overestimating with 2-3
m/s. Considering that a typical wind speed at Dome C in this season is
$\sim$3 m/s\cite{Hag08, Aris05} this corresponds to a large
discrepancy. The absolute temperature is in general warmer in the
ECMWF analyses than in the radiosoundings near the surface in winter,
achieving a difference of the order of $\sim$6 K. The wind speed and
the temperature show similar trends above the two sites with exception
of the absolute temperature in winter. At the South Pole the
temperature from the analyses appear warmer in the whole 150 m while
at Dome C only in the first 20 m.

These conclusions also confirm our doubt expressed in the Introduction
concerning the paper of Sadibekova et al. (2006)\cite{Sadib06} that
claimed a good agreement between ECMWF analyses and radiosoundings
also in the vicinity of the surface. 

In spite of the fact that their study was based on the ERA-reanalyses
(a product having a lower resolution than the MARS catalogue used in
this study) the agreement between radiosoundings and analyses in their
data matches well with our findings that predict a good agreement
between ECMWF analyses and radiosoundings in summer. Our analysis,
extended to winter, reveals that in this season the agreement is far
from being good and the sharp changes in wind speed and temperature
closest to the surface measured by the radiosoundings are not well
reconstructed by the ECMWF analyses.

\begin{figure}
\begin{centering}
\includegraphics[]{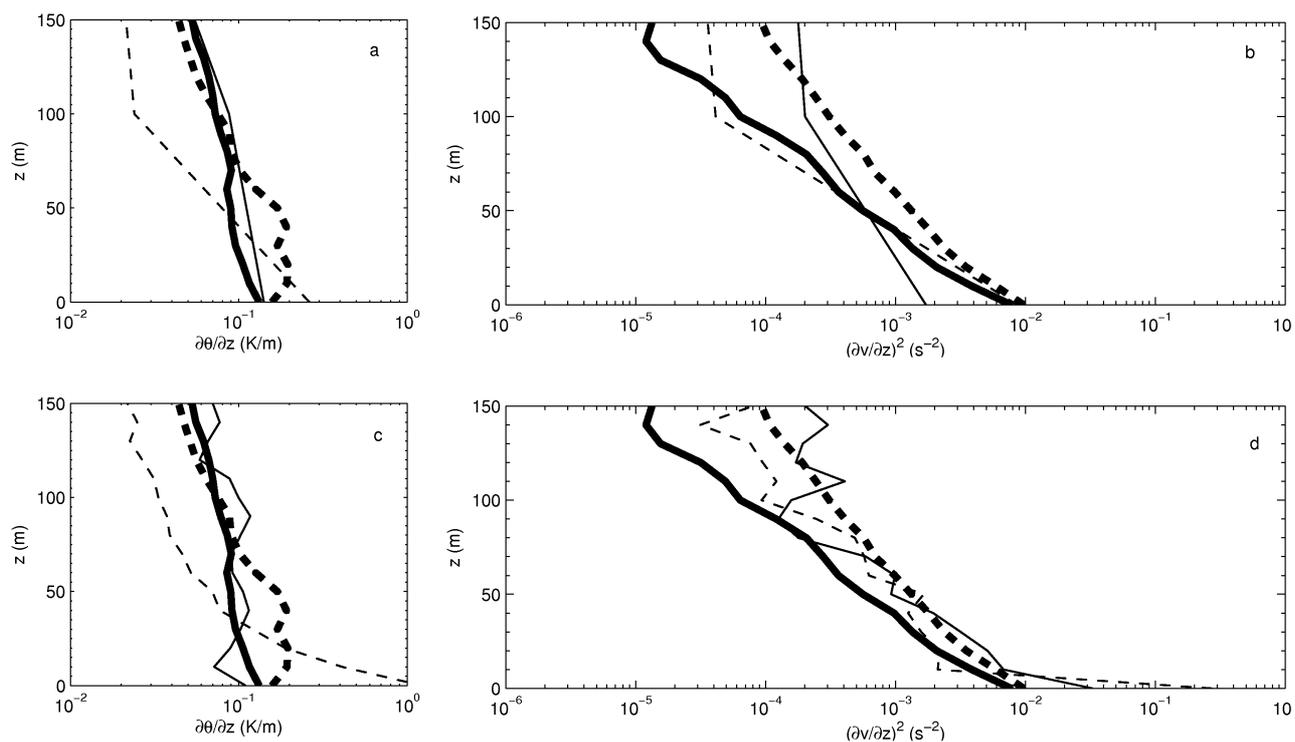}
\caption{The gradient of the potential temperature and square of the
  gradient of the wind speed near the surface for Dome C and the South
  Pole in July 2006. The dashed lines refer to Dome C and the solid
  lines refer to the South Pole. Thick lines are ECMWF analyses and
  thin lines are radiosoundings. In the top plots (a and b) the
  radiosoundings are interpolated with a step of 100 m, in the bottom
  plots (c and d) with a step of 10 m.
\label{gradis}}
\end{centering}
\end{figure}

In order to provide the most comprehensive and compact comparison of
ECMWF analyses and radiosoundings above Dome C and the South Pole
near the surface we prefer to change the focus to the two key
parameters that define the thermodynamic stability , i.e. the
gradients of the potential temperature and of the wind speed. Only a
study of the simultaneous systematic effects on both parameters can
tell us if it is possible to use ECMWF analyses to quantify the
thermodynamic stability in the surface layer.

Figure~\ref{gradis} shows the median gradient of the potential
temperature (left) and  the square of the gradient of the wind speed
(right) in the first 150 m with a vertical resolution of 100 m (a and
b) and of 10 m (c and d) in the radiosoundings. As expected the
radiosoundings show a sharper gradient than the analyses near the
surface. The ECMWF analyses are able to identify that the gradients
above Dome C are larger than those above the South Pole. Unfortunately
a precise quantification is not possible and, even in the case of the
best vertical resolution (c and d), the offset of the analyses with
respect to the radiosoundings of the two parameters
($\partial \theta/\partial z$ and $(\partial v/\partial z)^2$) is not
comparable above the two sites. This implies that the ECMWF analyses
do not smooth out the potential temperature and wind speed gradients
in a similar way above the two sites.

Knowing that the Richardson number depends on the ratio of
$\partial \theta/\partial z$ and $(\partial v/\partial z)^2$ we conclude
that it is pretty risky to draw any conclusions on a comparative
analysis of the Richardson number in the surface layer between
different sites calculated with the ECMWF analyses. As a consequence
it is possible to retrieve a ranking of the three sites with respect
to the thermal and the dynamic stabilities in an independent way, but it
is not possible to retrieve a ranking of the three sites with respect
to the Richardson number in the surface layer. In the free atmosphere
the ECMWF analyses are reliable and such a comparison will be
performed in Sec. 5.

\subsection{\boldmath$\partial \theta/\partial z$ and $(\partial v/\partial
  z)^2$ at the South Pole, Dome C, Dome A and Dome F}

As a consequence of the conclusions in the previous section, the
'thermal' and the 'dynamic' properties are shown independently. The
dynamic properties are described by the changes in wind speed with
height and the thermal stability is represented by the vertical
gradient of the potential temperature. A positive potential
temperature gradient is defined as stable condition, the vertical
displacement of air is suppressed and so is the production of dynamic
turbulence. 

\begin{figure} 
\begin{centering}
\includegraphics[angle=-90, scale=0.67]
{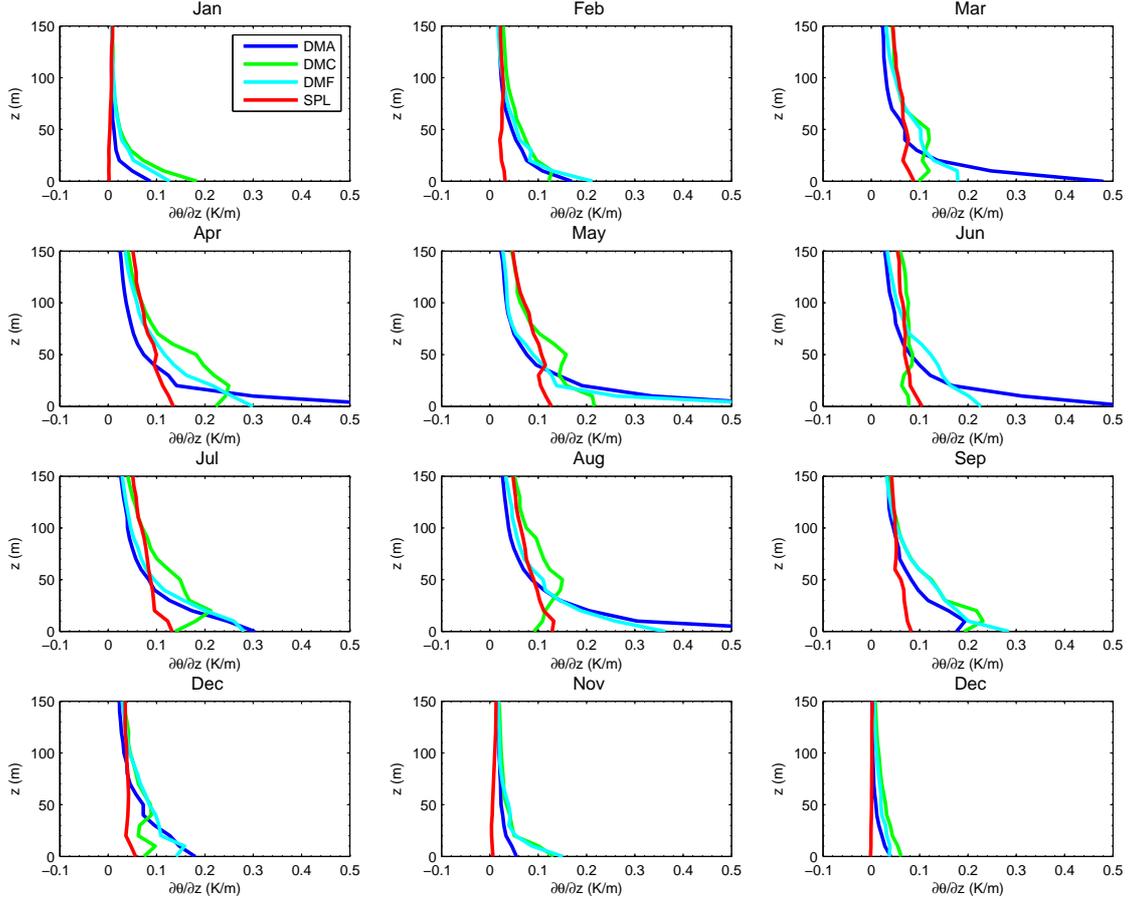}
\caption{The monthly median of the gradient of the potential
  temperature for 2005, Dome A (dark blue lines), Dome C (green
  lines), Dome F (light blue lines) and the South Pole (red
  lines).\label{dth}}
\end{centering}
\end{figure}

The absence of sunlight in the antarctic night and the consequent
radiative cooling of the snowy ice surface creates a strong
temperature inversion near the surface. The monthly median of the
gradient of the potential temperature in the first 150 m for the four
sites is shown in Fig.~\ref{dth}. The analyses data used for this
figure are the data from the most stable synoptic hour, i.e. the
synoptic hour closest to local midnight, 00 UTC at the South Pole and
Dome F and 18 UTC at Dome A and Dome C. These times corresponds to 23
LT (Local Time) for Dome A, 02 LT for Dome C and 03 LT for Dome F. An
inversion is always present during winter. In the central months of
the winter (June, July, August) Dome A (dark blue lines) shows the
most stable stratification, closely followed by Dome F (light blue
lines). The inversion above Dome A is from March to August clearly
larger than for the other sites and has a very sharp change in the
slope of the gradient at about 20 m above the surface. Dome C (green
lines) and the South Pole (red lines) also has a clearly stable
stratification but it is lesser than for the other two sites.

During the summer months the stratification is close to neutral,
particularly in December. When the stratification is neutral
($\partial \theta/\partial z \approx 0$) the vertical motion of air is
not suppressed but neither is it encouraged and a small perturbation
can trigger dynamic turbulence.  However, in this season there is a
diurnal variation in the antarctic boundary layer where the
stratification is close to neutral during the day and during the night
an inversion is created. There is a clear difference between the South
Pole, that during all summer has a neutral gradient while the domes
all show a clear temperature inversion near the surface during the
night, as shown in Fig.~\ref{dth}. To draw any
conclusion of the differences between the domes is quite risky in this
season as the temporal resolution is a little too low.

For the median of the gradient of the wind speed in the first 50 m
above the surface of the South Pole, Dome C and Dome A we refer the
reader to Fig.~5 of Hagelin et al (2008)\cite{Hag08}. The low
resolution of the ECMWF analyses makes it difficult to see any clear
differences between the sites for this parameter, but almost every month the gradient
is largest at the lowest level for Dome A.

\section{RADIOSOUNDINGS: THE SURFACE WIND SPEED}

\begin{figure}
\begin{centering}
\includegraphics[angle=-90, scale=0.5]
{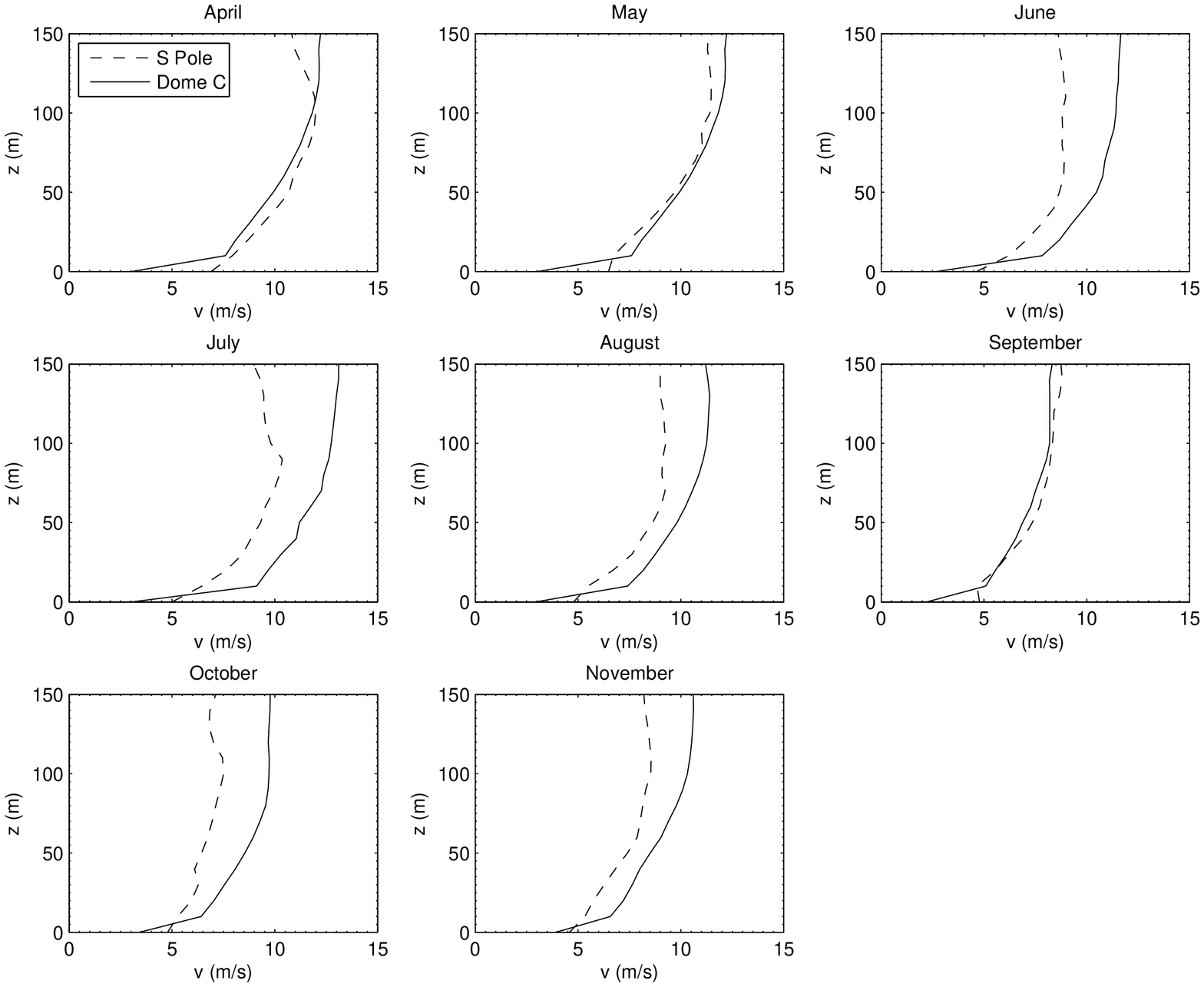}
\caption{The wind speed near the ground at Dome C (solid lines) and the
  South Pole (dashed lines), April to November 2006.\label{rs2}}

\includegraphics[angle=-90, scale=0.6]
{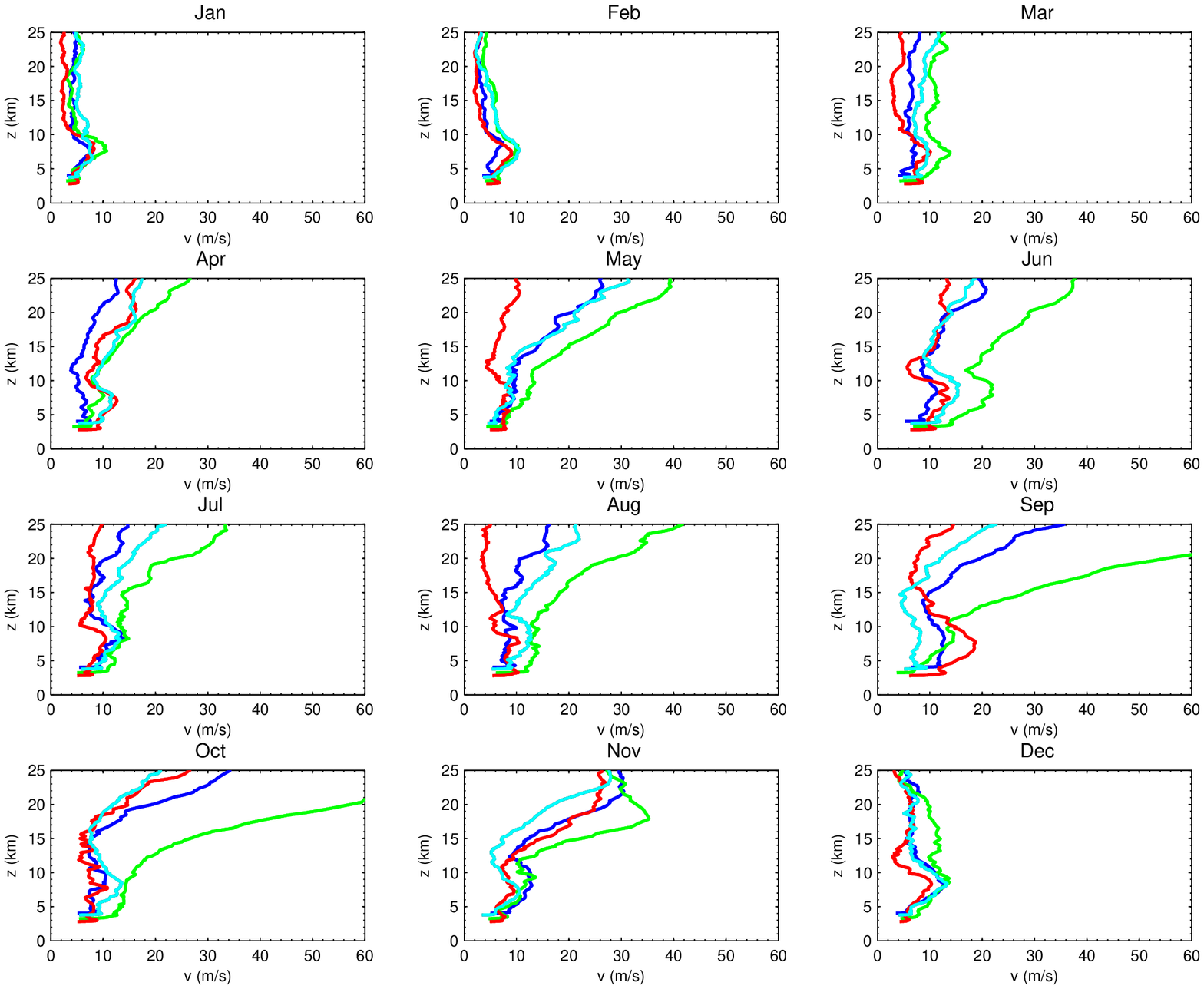}
\caption{The monthly median wind speed for 2005 00 UTC for Dome A
  (dark blue lines), Dome C (green line), Dome F (light blue) and the
  South Pole (red lines). \label{wi}}
\end{centering}
\end{figure}

Above the summits of the Internal Antarctic Plateau the surface winds
are expected to be weaker than elsewhere on the
Plateau. Fig.~\ref{rs2} shows the median wind speed near the surface
measured with radiosoundings at the South Pole (dashed lines) and Dome
C (solid lines) from April to November. While it is true that the wind
speed at the lowest level is weaker at the summit (Dome C) than at the
slope (South Pole), there is a sharp increase in the wind speed above
Dome C in the first few tens of meters. At the height of 10/20 m, from
May to November i. e. winter, the wind speed is higher above Dome C
than above the South Pole. Above this height the wind speed at Dome C
is either higher than or very similar to the wind speed at the
South Pole.

In the core of the winter (June, July, August) the wind speed above
Dome C reaches 8 m/s at 20 m and 9 m/s at 30 m. The sharp change in
wind speed in the first 10/20 m matches our expectations of a large
wind speed gradient. This is a necessary condition to justify the
presence of optical turbulence in the surface layer\cite{Agabi06} in
spite of very stable thermal conditions (i.e. a positive gradient of
the potential temperature). 

Trinquet et al (2008)\cite{Trinq08}, using a small sample of
radiosoundings in winter, observed a wind speed of 5 m/s at 20 m
altitude and only 8 m/s at 70 m. Our result, obtained with a complete
statistical sample in winter, tells us that their estimate is too
optimistic and we should expect a larger wind speed at low altitudes.

Mechanical vibrations in the first 10/20 m, that might be derived from
the impact of the atmospheric flow, flowing at 8-9 m/s, on a telescope
structure, are probably more critical above Dome C than above the
South Pole. This should be taken into account carefully when designing
astronomical facilities.

\section{ECMWF ANALYSES: THE WEAKEST WIND SPEED IN THE TROPOSPHERE AND
  THE LOWER STRATOSPHERE}

The monthly median of the wind speed, from the ground up to 25 km
a.s.l. for Dome A (dark blue line), Dome C (green line), Dome F (light
blue line) and the South Pole (red line) is shown in
Fig.~\ref{wi}. During summer the wind speed is quite weak in the whole
height interval for all sites. A small maximum is observed at roughly
8 km a.s.l. From December to March the median wind speed is never
larger than 15 m/s at any height for any site. As the winter
approaches the wind speed in the upper atmosphere increases. Above 10
km a.s.l. the wind speed increases monotonically and the highest wind
speed is more probably found in the higher part of the atmosphere. The
rate at which the wind speed increases is not the same above the
different sites. The smallest increase rate is observed at the South
Pole whereas the maximum rate is found above Dome C. These differences
are far from being negligible since at these heights the median wind
speed at Dome C is almost twice that of the other three
sites. Geissler \& Masciadri (2006)\cite{Ges06} found a similar trend
of the wind speed at Dome C for the years 2003 and 2004. The trend of
the wind speed is therefore confirmed in different years.

The difference in the slope of the vertical profile of the wind speed
in Fig.~\ref{wi} is most likely related to the special synoptic
circulation over Antarctica. The jet stream, that characterize the
vertical wind speed profile at mid-latitude sites at the tropopause
level, does not form over Antarctica. The large scale circulation is
instead dominated by the polar vortex which creates strong
high-altitude winds surrounding the polar high in
winter\cite{Schwe84}. The wind speed increases monotonically above 10
km (the height of the tropopause). The result observed in
Fig.~\ref{wi} can be explained by the fact that the South Pole is
located near the polar high and consequently the influence of the
polar vortex is weak at this site. The domes are situated further from
the centre of the continent, and thus from the centre of the polar
vortex. It is to be expected that the wind speed is larger above these
sites. The farther from the polar high (the centre of the polar
vortex) the site is, the greater is the strength of the wind speed
above the tropopause.

The assumption that the strength of the wind speed in the upper
atmosphere is correlated to the distance of the site from the centre
of the polar vortex means that if we knew the exact position of the
polar high we would have an excellent tool to identify, \emph{a
  priori}, the site site with the weakest wind speed above the
tropopause during the winter. Unfortunately the polar vortex is not a
perfect cone so the centre of the vortex is not located at the same
coordinates at different heights. Looking at Fig.~\ref{wi}, it is
evident that the South Pole and Dome C respectively have the weakest
and the greatest wind speed above the tropopause. The wind speed above
Dome A and Dome F is mostly comparable and it is more difficult to
tell which site has the weaker wind speed. The polar high probably
fluctuates in a region located between the South Pole, Dome A and Dome
F. (See Fig.~8 in Hagelin et al. (2008)\cite{Hag08})


\section{THE RICHARDSON NUMBER AT THE SOUTH POLE, DOME C AND DOME A}

The Richardson number is an indicator of the stability of the
atmosphere,
\begin{equation}
Ri=\frac{g}{\theta }\frac{\partial \theta/ \partial z}{(\partial v/ \partial
  z)^2}
\end{equation}
where g is the gravitational acceleration (9.8 m/s$^2$), $\theta$ is
the potential temperature and v is the horizontal wind speed. The
atmosphere is considered to be stable if the Richardson number is
larger than a critical value, typically 0.25. If the Richardson number
is less than the critical number the stratification is classified as
unstable. The smaller the Richardson number is the higher is the
probability of the triggering of turbulence.

\begin{figure}
\begin{centering}
\includegraphics[angle=-90,scale=0.55]
{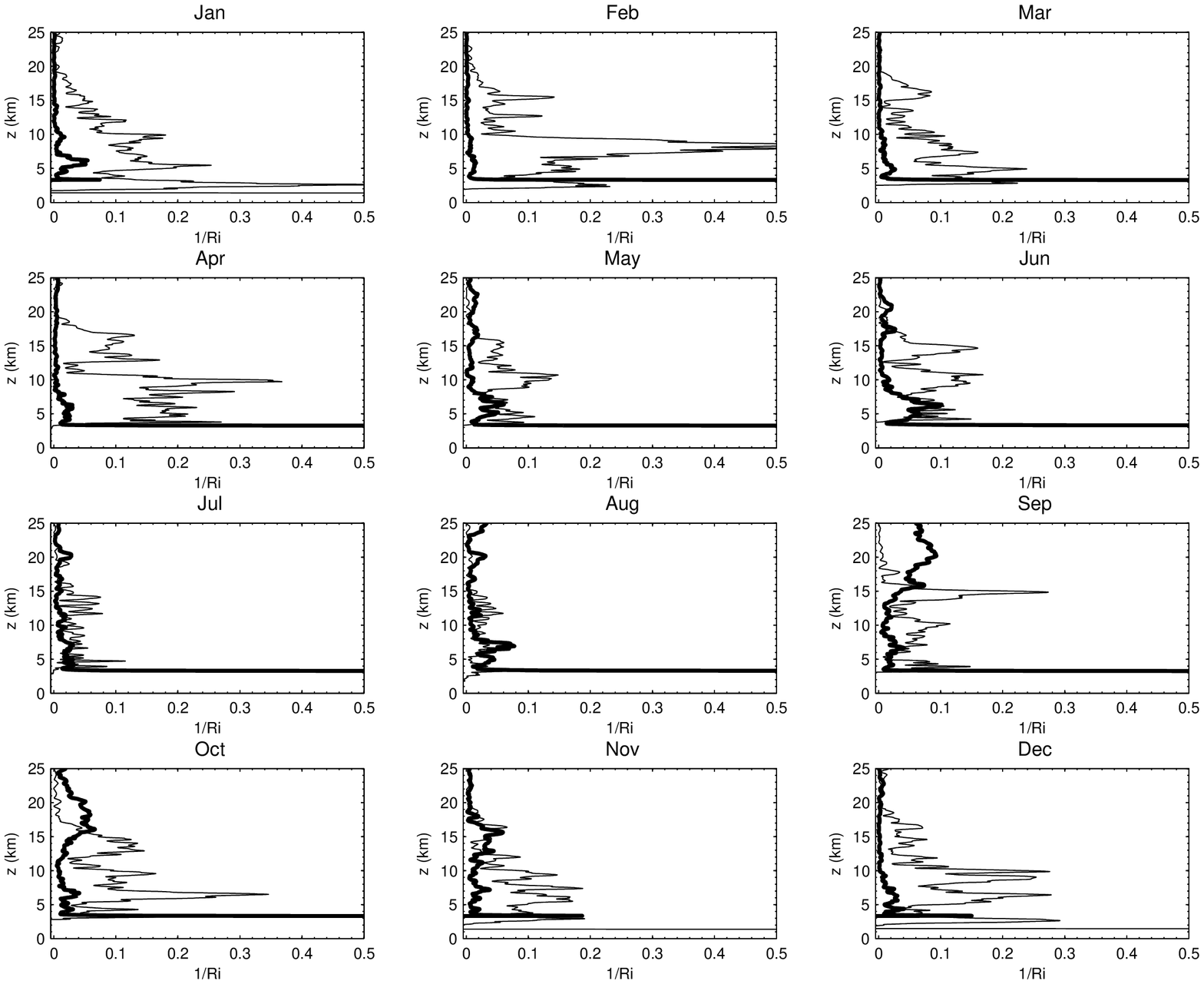}
\caption{The monthly median for 2005 of the inverse of the Richardson number
  (1/Ri) for Dome C (thick lines) and Mt. Graham (thin lines).\label{iRi1}}

\includegraphics[angle=-90,scale=0.55]
{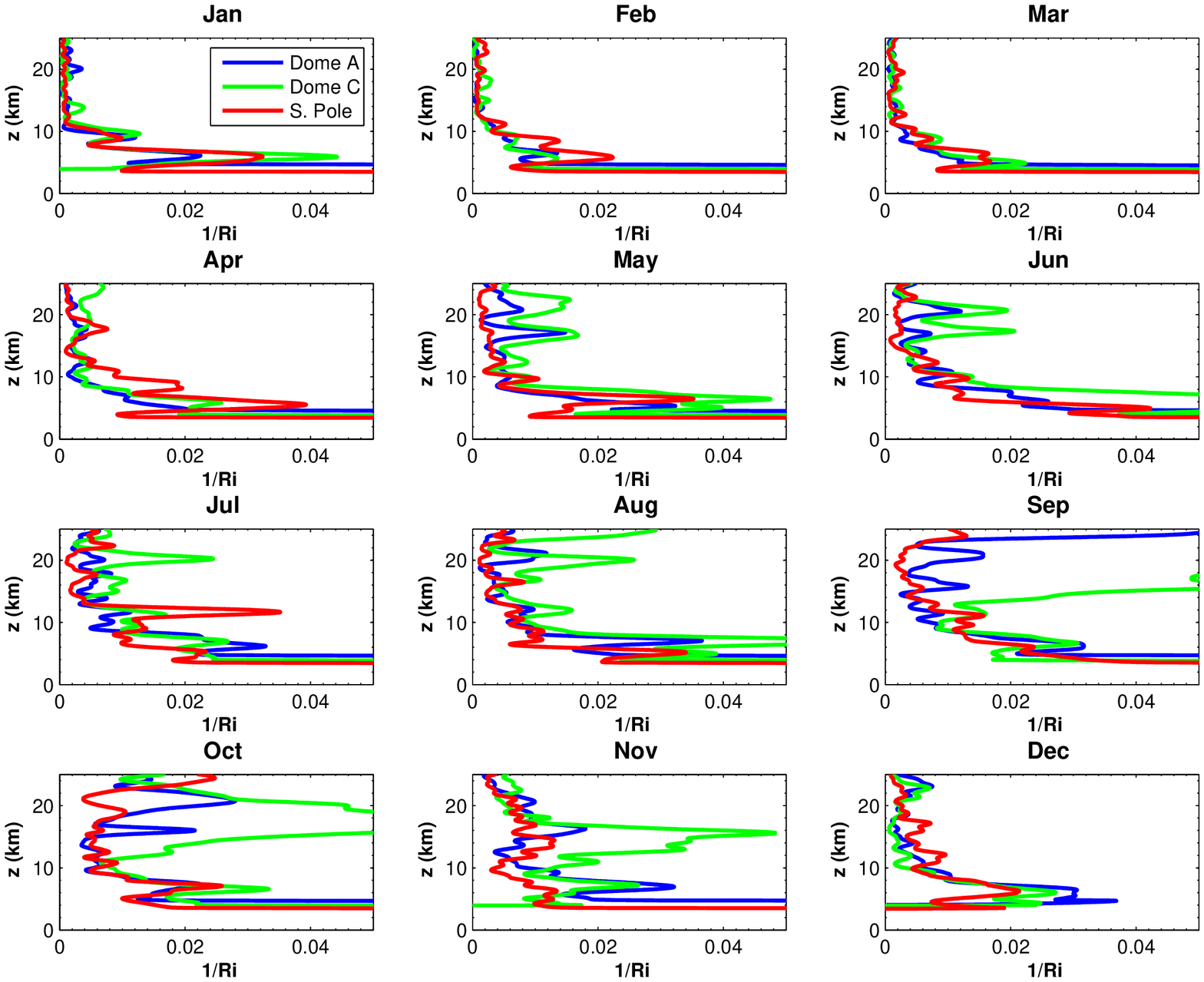} 
\caption{The monthly median for 2005 of the inverse of the Richardson number
  (1/Ri) for Dome A (blue lines), Dome C (green lines) and the South Pole
  (red lines).\label{iRi2}}
\end{centering}
\end{figure}

Comparing the Richardson number gives a relative estimate of the
probability of the triggering of turbulence above different sites.
Fig.~14 of Geissler \& Masciadri (2006)\cite{Ges06} shows the median of
the inverse of the Richardson number (1/Ri) for Dome C in the first 20
km of the atmosphere. However such a result can only give a
comparative analysis that necessarily is qualitative. It is possible
to identify if one region shows a higher or smaller probability to
trigger turbulence than another region but we do not have a reference
to quantify these differences nor can we tell whether these differences
are negligible or not. In order to investigate this idea further we
calculated (Fig.~\ref{iRi1}) the median of 1/Ri from Dome C (thick
lines) and from Mt Graham, Arizona (thin lines), taken as an example
of a typical mid-latitude site, for each month during 2005. The
inverse of the Richardson is shown instead of the Richardson number
because the inverse shows a better dynamic. The smaller 1/Ri is, the
more stable is the atmosphere and the 1/Ri is smaller above Dome C
than above Mt Graham almost everywhere. This result definitely confirms
the method proposed by Geissler \& Masciadri (2006)\cite{Ges06} to
use the 1/Ri as a tool to rank different sites with respect
to the probability to trigger turbulence. During September and
October, when the median wind speed at high altitudes is remarkably
strong at Dome C (see Fig.~\ref{wi}), 1/Ri is larger than for the
mid-latitude site. Consequently this region needs to be carefully
monitored as the strong increase in wind speed increases the 1/Ri and
thus the probability to trigger instabilities is larger or comparable
to that over a mid-latitude site.

The vertical median profiles of 1/Ri for the three Antarctic sites
(South Pole, Dome C and Dome A) are shown in Fig.~\ref{iRi2}. During
the summer the profiles from the different sites are very similar to
each other. The 1/Ri has a maximum at ground level and a smaller peak
somewhere slightly above 6 km a.s.l. and above $\sim$10 km the
atmosphere is very stable. From April/May the instability above 10 km
increases. At the end of winter (September/October) the instability in
the upper part of the atmosphere is even larger than the peak near 6
km for Dome A and Dome C. The instability above Dome C is more
pronounced than that of Dome A. Note that the wind speed is also
larger at Dome C in this region (Fig.~\ref{wi}). Above the South Pole
1/Ri shows the best conditions (i.e. the atmosphere is most stable)
over the whole year.

\section{CONCLUSIONS}

From a first comparison of the atmospheric characteristics of three
different sites on the Internal Antarctic Plateau: Dome A, Dome C and
the South Pole using ECMWF analyses and radiosoundings we can draw the
following conclusions.

The comparison of the ECMWF data to the radiosoundings shows that the
ECMWF analyses are reliable above the Internal Antarctic Plateau
during all seasons and altitudes except in the first tens of
meters. During no season does the difference in wind speed exceed 1
m/s above this height. The median difference in absolute temperature
is within 2 K in the same range. The discrepancy of the wind speed 
in the surface layer is slightly larger, 2-3 m/s, while the
analyses also show a tendency to overestimate the temperature in the
lowest level in winter by $\sim$4-5 K. However, one should also
remember that during winter the radiosoundings seldom reach an
altitude higher than 10-12 km, making it difficult to estimate the
reliability of the analysis at high altitudes during the cold season.

Near the surface the estimate of the ECMWF analyses are less accurate
and this justify the use of mesoscale models when characterizing the
boundary layer. However it is also fundamental to verify that the
mesoscale models do provide a better estimate than the General
Circulation Models (GCM) in this layer. 

The gradient of the potential temperature indicates a stable
stratification at all four sites, even if the ECMWF analyses
generally underestimate this gradient when compared to measurements
from radiosoundings. The stability is particularly strong in winter
where Dome A is the site with the steepest gradient, followed by Dome
F. The site with the least stable conditions, in this study, is the
South Pole.

The median wind speed close to the surface is weaker at Dome C than at
the South Pole during the period investigated (April to
November). However, the wind shear at Dome C is much larger than that
of the South Pole and so that at 10/20 m the wind speed of Dome C is
comparable or larger to the South Pole. Such a strong wind shear in
combination with a stable thermal stratification in this layer is most
likely the cause of the intense optical turbulence measured in the
first tens of meters at Dome C\cite{Agabi06}. Such a strong wind speed,
$\sim$8-9 m/s at 10 m, might be a source of vibrations produced by the
impact of the atmospheric flow on telescope structures and should
therefore be taken into account in the design of astronomic
facilities.

In the free atmosphere above 10 km, the polar vortex causes
the wind speed to increase monotonically in winter. The increase rate
is proportional to the distance of the site from the polar high. This is
the reason why Dome C shows the largest wind speed above 10 km in
winter. At 15 km altitude the wind speed of Dome C can easily be
almost twice of the wind speed of Dome A or Dome F and even thrice to
that of the South Pole. The wind speed above the South Pole is the
weakest of the four in all seasons and at all heights. 

The monthly median of the inverse of the Richardson number (1/Ri)
indicate that the probability to trigger turbulence is larger above a
mid-latitude site (for which we have a reliable characterization of
the turbulence) than above any site investigated by us at the Internal
Antarctic Plateau, except the upper atmosphere at Dome C during
September and October. This is a definite proof that the method
proposed by Geissler \& Masciadri (2006)\cite{Ges06} is
reliable. Moreover, using this method to compare our three antarctic
sites the 1/Ri-values indicated that Dome A and the South Pole show
more stable conditions than Dome C above the first 100 m. This is
probably due to the polar vortex that increases the wind speed in the
upper atmosphere, which also increases the probability to trigger
turbulence. Near the surface it is risky to use estimates of 1/Ri from
the ECMWF data as we do not observe an equivalent smoothing effect for
the gradients of potential temperature and wind speed.

This study allowed us to draw a first comprehensive picture of the
atmospheric properties above the Internal Antarctic Plateau. In spite
of the generally good conditions for astronomical applications, Dome
C does not appear to be the best site with respect to the wind speed,
in the free atmosphere as well as in the surface layer. All the other
three sites show a weaker wind speed in the free atmosphere. The
estimates of the surface layer needs to be considered with some caution as
the ECMWF analyses are not optimized for this layer and radiosoundings
are only available for Dome C and the South Pole. Above Dome A the
gradient of the potential temperature is particularly large close to
the surface, indicating extreme thermal stability associated to a
strong value of the optical turbulence in this layer when a
thermodynamic instability occurs. It is even possible that the optical
turbulence here is larger than at Dome C. However, our study showed
that, to predict the thickness of such a layer it is necessary to have
either measurement or simulations made with a mesoscale model
with a higher spatial resolution. This is a part of our forthcoming
activities.

At present, the real solid argument that makes Dome C a better place
for astronomical applications than the South Pole is the extreme
thinness of the optical turbulence surface layer. Dome A probably has
comparable or larger optical turbulence values with respect to Dome C
in the surface layer. We cannot conclude whether the surface layer at
Dome A is thinner than that observed at Dome C. However our study
clearly indicates that Dome C is not the best site on the Internal
Antarctic Plateau with respect to the wind speed (both in the surface layer
and in the free atmosphere) nor is it the site
with the most stable conditions in the free atmosphere. Both Dome A
and the South Pole show more stable conditions.

\acknowledgments     
 
This study has been carried out using radiosoundings from the AMRC
(Antarctic Meteorological Research Center, University of Wisconsin,
Madison $ftp://amrc.ssec.wisc.edu/pub/southpole/radiosonde$) and from
the Progetto di Ricerca 'Osservatorio Meteo Climatologico' of the
Programma Nazionale di Ricerche in Antartide (PNRA),
$http://www.climantartide.it$. ECMWF products are extracted from the
Catalogue MARS, $http://www.ecmwf.int$.  This study has been funded by
the Marie Curie Excellence Grant (FOROT) - MEXT-CT-2005-023878.


\bibliography{myref}  
\bibliographystyle{spiebib}   

\end{document}